\documentclass[%
 reprint,
showpacs,preprintnumbers,showkeys,
 amsmath,amssymb,
 aps,
]{revtex4-1}

\usepackage{graphicx}% Include figure files
\usepackage{dcolumn}% Align table columns on decimal point
\usepackage{bm}% bold math
\newcommand{\be}{\begin{equation}}
\newcommand{\ee}{\end{equation}}

\begin{document}

\title{Structures in the energy distribution of the scission neutrons:
finite neutron-number effect}

\author{N. Carjan$^{1,2,3}$} 
\email{carjan@theory.nipne.ro} 
\author{M. Rizea$^1$}
\affiliation{$^1${``}Horia Hulubei{"} National Institute of Physics and
Nuclear Engineering, Bucharest-Magurele, RO-077125, Romania
\\$^2$Joint Institute for Nuclear Research, FLNR,
141980 Dubna, Moscow Region, Russia
\\$^3$ CENBG, University of Bordeaux, 33175 Gradignan, France}

\date{\today}% It is always \today, today,
\begin{abstract}

{The scission neutron kinetic energy spectrum is calculated for $^{236}U$ 
in the frame of the dynamical scission model.  
The bi-dimensional time dependent Schr\"{o}dinger equation with time dependent 
potential is used to propagate each neutron wave function during the scission 
process which is supposed to last $1\times 10^{-22}$ sec.  
At the end, we separate the unbound parts and continue to propagate them as 
long as possible (in this case $50\times 10^{-22}$ sec) in the frozen fragments 
approximation. At several time intervals, the Fourier transforms of these wave 
packets are calculated in order to obtain the corresponding momentum 
distributions which lead to the kinetic energy distributions.
The evolution of these distributions in time provides an interesting insight 
into the separation of each neutron from the fissioning system and 
asymptotically gives the kinetic energy spectrum of that particular neutron.
We group the results in substates with given projection $\Omega$ of the 
angular momentum on the fission axis to study its influence on the spectrum.
Finally, the sum over all $\Omega$ values is compared with a typical 
evaporation spectrum as well as with recent precise measurements in the 
reaction $^{235}U(n_{th},f)$. Structures are present both in the 
scission-neutron spectrum and in the data.
}
\end{abstract}
%\pacs{02.30.Nw; 02.60.Lj; 02.70.Bf; 21.10.Pc; 25.60.Gc; 25.85.w}
%Use showkeys class option if keyword
\maketitle
\section{Introduction}
\label{intro}

It is now generally accepted that prompt fission neutrons (PFN) have two 
components with unknown relative intensities. In chronological order, these 
components are: neutrons dynamically released at scission (SN) and neutrons 
evaporated from fully accelerated fragments (EVN).
There is no indication which of these two components is the dominant one 
since the gross features of PFN can be reproduced by both models 
\cite{PLB,PROMA,CRT-ND16,CCC,Lemaire,Vogt,Serot,Kawano,GEF,Capote}.
To determine the relative percentage of SN and EVN, instead of looking at 
averaged properties, one has 
to analyze PFN observables correlated with fragment properties
in order to remove the above mentioned ambiguity.   

It is also important to find differences, even small, between the predicted
properties of 
the scission and evaporated neutrons that may be investigated experimentally,
thus making the separation of the two components possible.
It has been already pointed out \cite{CRT-ND16} that, for a fixed 
fragment-mass division,
the angular distributions with respect to the fission axis of EVN and of SN 
are different: the first is smooth while the second presents oscillations
due to the proximity of the fragments at the moment of emission. 

This time we concentrate on the kinetic energy spectrum of the 
scission neutrons, again for a given fragment-mass ratio. We calculate it 
for neutrons with quantum numbers 
$\Omega$ = 1/2, 3/2, 5/2, 7/2 and 9/2. $\Omega$ is the projection of the 
angular momentum on the fission axis. 
They account for  99$\%$ of the total multiplicity. The result 
is compared with a typical evaporation spectrum to reveal differences.       

Sec 2 contains the description of the model used. The corresponding equations
are given in Sec 3. Numerical results for individual neutron states in 
$^{236}U$ are presented in Sec 4. In Sec 5 the total energy spectrum is 
calculated and compared with recent measurements.
The summary is in Sec 6. 

\section{From bound to free neutrons} 
\label{sec-2}

In order to calculate the kinetic energy spectrum of the scission neutrons 
we need to identify the part of each neutron wave packet which left the 
fissioning system and therefore represents a free neutron.  

We do this in the frame of the dynamical scission model \cite{RCNPA13} 
in which the fissioning system 
undergoes a diabatic transition during the neck rupture. Due to the coupling  
with the rapidly changing potential, each initially bound neutron state becomes 
a wave packet with few components in the continuum. This process is simulated 
introducing a time-dependent potential (TDP) in the two-dimensional 
time-dependent Schr\"{o}dinger equation (TDSE2D).
The model is best suited to low energy fission: spontaneous or sub-barrier. 
An amount of excitation energy at the last saddle point could lead 
to a neutron evaporation before scission which is 
not included in the present calculations. 

There are three parameters in the dynamical scission model: the nuclear shapes 
just before ($\alpha_i$) and immediatelly after scission ($\alpha_f$) and 
the duration $\Delta T$ of the transition between these two shapes. 
These quantities are not really known; one can only make educated guesses about 
them. The lower limit of $\Delta T$ should be
about $5\times 10^{-23}$ sec i.e., the time required for a Fermi level nucleon to
cross a 4 fm thick neck. A value of $\Delta T$ between 1 and 2 $\times 10^{-22}$ sec 
can therefore be considered realistic.
The minimum neck radius $r_{min}$ in the initial configuration predicted
by the optimal scission shapes \cite{Fedo} is $\approx$ 2 fm. 
It is a generally accepted value since 
it can be deduced also from general considerations like the size of the
alpha particle. We take a slightly lower value (1.6 fm).
There is no indication about the  minimum distance between the surfaces of the
two fragments $d_{min}$ in the final configuration.  We take 0.6 fm.
These $r_{min}$ and $d_{min}$ values were already used in our first publication 
\cite{CTS}  and have never been changed. They lead to 
an average scission neutron multiplicity of 0.6 neutrons per fission event, 
i.e., to only 25$\%$ of the total prompt fission neutron multiplicity.
Although we know \cite{PLB} that both 
$<\nu_{sc}>$ and the average kinetic energy $<E_{kin}>$ are sensitive to the 
the parameters of the model, we do not think that it makes 
sense to adjust them to the existing experimental values for all prompt 
fission neutrons (i.e., to 2.41 and 1.99 MeV respectively, obtained in the 
reaction $^{235}U(n_{th},f)$). 
When more reliable values for these quantities are available we will use them  
and find out how significant is the percentage of neutrons released at scission 
and emitted during the acceleration of the fragments.
In fact, self-consistent microscopic 
models, such as the density functional 
theory extended to superfluid systems and real-time dynamics \cite{Bulgac},
could provide estimates for the three parameters of our model. 

 The unbound components of the neutron wave packet will start  
leaving the nascent fragments immediately after scission but this separation 
takes time. 
Hence they leave during the acceleration phase: up to approximately 
$6\times10^{-21}$ sec for most of them . This is a rough estimation 
based on the half-live of neutron emission at scission which is about 
$2\times10^{-21}$ sec \cite{RCNPA13} if $\Omega$=1/2. 
Large times require large spatial grids. 
Although we implemented transparent boundary conditions \cite{Hadley}, 
the reflexions on the boundaries of the numerical grid are not completely 
reduced and we need to push our computational resources to their limit.

At the beginning, i.e., immediately after scission, the unbound neutrons are 
mainly localized inside the nucleus and therefore possess very high kinetic 
energies (of the order of the depth of the potential). To obtain the measured 
spectrum, one has to wait until these neutrons are outside the fissioning 
system. This detachment is simulated with 
TDSE2D, using a constant potential this time. 
We stop at $T_{max}= 5\times10^{-21}$ sec when the percentage of unbound 
neutrons that are still inside the nucleus attains a minimum (about 10$\%$).    

Since the neutron motion is much faster than the separation of the nascent 
fragments, the freeze of the fissioning nucleus at its configuration 
immediately after scission is justified and it simplifies our numerical task.   
Even when the neutrons are outside the fragments, their kinetic energy is at 
least 1.5 MeV (see Figs. 1 to 8). The total kinetic energy of the fully 
accelerated fragments is 0.75 MeV/nucleon on the average.
Therefore, at the begining of the acceleration phase 
when the scission neutrons are emitted, the velocity of the fragments is 
negligible as compared with the velocity of the neutrons.    

The Fourier transforms of the unbound-neutron wave packets give, at each time, 
the momentum distributions and therefore also the kinetic energy distributions.
Asymptotically, the sum over all neutrons, weighted with their occupation 
probabilities, leads to the scission neutron spectrum. 

Let us now put the description from above into equations.
\section{Formalism}
\label{sec-3}

The scission consists in the neck rupture and the absorption of the neck stubs 
by the nascent fragments.
We consider axially symmetric fissioning nuclei and use cylindrical coordinates.
Let $|\Psi^i(\rho,z)\rangle$ be the eigenfunctions
of the Hamiltonian of independent neutrons in the just-before-scission 
configuration.
During the scission process these functions evolve in a time-dependent 
potential according to TDSE2D: 
\be
\label{bid0}
i \hbar \frac{\partial \Psi^i(\rho,z,t)} {\partial t} =
{\cal{H}}(\rho,z,t) \Psi^i(\rho,z,t).
\ee
The solution is obtained using a numerical scheme of Crank Nicolson type
\cite{riz08,CICP}. The infinite physical domain is replaced by a finite grid:
$[0,\rho_{max}]\times [-z_{maz},z_{max}] = [0,84 fm]\times [-128 fm,128 fm]$ 
with 
$\Delta \rho = \Delta z = 1/8$ fm.  For the time evolution we use a step 
$\Delta t = 1/128\times 10^{-22}$ sec. Special conditions on the boundaries 
of the grid are imposed to reduce reflexions \cite{Hadley}.

In the non-adiabatic regime, the propagated wave functions 
$|\Psi^i(\rho,z,t)\rangle$ are wave packets which have also positive-energy 
components.

The probability amplitude
that a neutron occupying the state $|\Psi^i\rangle$ before scission
populates an eigenstate $|\Psi^f\rangle$ immediately-after scission is
\begin{widetext}
\be
a_{if} = \langle\Psi^i(\Delta T)|\Psi^f\rangle =
2\pi \int \int (f_1^i (\Delta T) f_1^f + f_2^i (\Delta T) f_2^f) \rho d\rho dz.
\ee
\end{widetext}
$a_{if}$ is $\neq$ 0 only if $|\Psi^i\rangle$ and 
$|\Psi^f\rangle$ have the same
projection $\Omega$ of the total angular momentum. 
$\Delta T$ is the duration of the scission process assumed here to be 
$10^{-22}$ sec i.e., relatively short. $f_1$ and $f_2$ are
the two components of the wave function corresponding to spin up and spin 
down respectively.

The probability that this neutron is unbound at the end of the scission 
process is given by:
\be
P_{em}^i = v_i^2(\sum_{unbound}|a_{if}|^2) =
v_i^2 (1 - \sum_{bound}|a_{if}|^2)
\ee
where $v_i^2$ is its initial occupation probability.

The part of the wave packet which is in the continuum at  $\Delta T$:
\be
|\Psi^i_{em}\rangle = |\Psi^i(\Delta T)\rangle - \sum_{bound} a_{if}
|\Psi^f\rangle.
\ee
will leave the fissioning nucleus and asymptotically will describe the 
emitted scission neutron. 

To calculate the scission neutron spectrum we have therefore to propagate 
$|\Psi^i_{em}\rangle$ for as long as possible, let's say until 
$\Delta T +T_{max}$ with $T_{max} = 50\times 10^{-22}$ sec.  
Since the separation of the 
fragments is slower than the neutron emission, for the sake of simplicity, 
we keep the fragments in their configuration at $\Delta T$. 

In order to vizualize the detachment of the unbound fractions of the neutron 
wave packets
from the fissioning system, we extract at several times $\Delta T +T$ these
fractions and calculate their Fourier transform \cite{EPJA,RCT4}:
\begin{widetext}
\be
\label{eq5}
F^i(k_{\rho},k_z,T) =  2 \pi\int_{-\infty}^{\infty} \left[\int_{0}^{\infty} 
\Psi^i_{em}(\rho,z,T)
J_0(2\pi \rho k_{\rho}) \rho \mathrm{d}\rho \right] \mathrm{e}^{-2 \pi 
\mathrm{i} z k_z} \mathrm{d}z
\ee
\end{widetext}

In this way we can study the probabilities both in coordinate and in 
momentum space as a function of time.
$J_0$ is the zero-order Bessel function of the first kind. The 
transform with respect to the variables $\rho,k_{\rho}$ is called the 
zero-order Hankel transform. Thus, the Fourier transform in cylindrical 
coordinates implies a combination of Hankel and one-dimensional Fourier 
transforms. 
The present study represents the first application in nuclear physics of such 
transforms.

\section{Post-scission evolution of the unbound neutrons and of their 
kinetic energies}
\label{sec-4}

Calculations are performed for the fission of $^{236}U$ having in mind 
the reaction $^{235}U(n_{th},f)$ which has been re-measured recently with
better statistics and
improved resolutions in mass, angle and energy \cite{Alf,AlfT4}.
The pre- and post-scission nuclear shapes are described by Cassini ovals 
\cite{PAS71} with only two parameters corresponding to the overall elongation 
and the mass asymmetry \cite{CR2010}.  
Numerical results for the most probable mass division (light fragment 
mass $A_L=96$) are presented.
 
We have calculated the Fourier transform using Eq. (5) for 
wavefunctions corresponding to $\Omega=1/2, 3/2, 5/2$, $7/2$ and $9/2$. 
Each point in the ($k_{\rho},k_z$) plane corresponds to an absolute value 
$K=\sqrt{k_{\rho}^2+k_z^2}$ and a probability 
$P=|F(k_{\rho},k_z)|^2 k_{\rho}\Delta k_{\rho}\Delta k_z$
that a scission neutron has its momentum $\vec{K}$ in the volume element 
$\mathrm{d}^3\vec{K}$. The points of constant K-value lie on a circle.
Since the Fourier transform is given only on the grid points we can represent 
the $K$-distribution only as a histogram.
For this we divide the domain of $K$-values in equal intervals and
group the grid points according to the interval to which they belong. Summing
up the probabilities of the points in each group one obtains the
probability $P_i(K)$ that a given neutron $i$ has its $K$-value in the 
respective interval.
From the momentum distribution one can deduce the kinetic energy distribution,
$P_i(E_{kin})$, using the relation $E=\frac{\hbar^2}{2\mu} K^2$ and multiplying 
with the Jacobian $dE/dK \sim E^{1/2}$ in order to accommodate for this change 
of variable.

In the figures 1 to 3 are shown unbound wave packets
for $\Omega$=1/2 and indices $i$ = 22, 26 and 28 
(as sum of square moduli of the two components $f_1$ and $f_2$) juxtaposed 
with kinetic-energy histograms at different times $T$ after scission.

\begin{figure*}[!htbp]
\centering
\includegraphics[width=0.80\textwidth,clip]{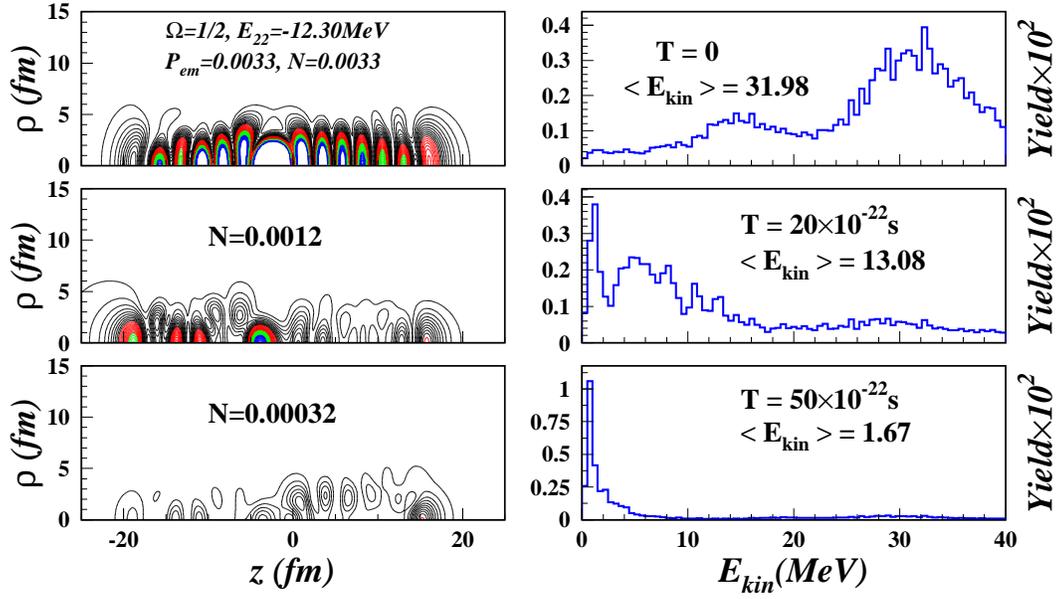}
\caption{Square modulus of the unbound WF$_{22}$ (left column) and energy 
distribution (right column) at 
different times T. The wavefunctions at $T=0$ and $50\times 10^{-22}$ sec 
are represented relative to that at $T=20\times 10^{-22}$ sec. 
The values on the ordinates of the histograms are $P_{22}(E_{kin})$ 
probabilities multiplied by $10^2$. 
$E_{kin}^{mean}= \frac{\sum_{m,n} E_{kin} P E_{kin}^{1/2}}
{\sum_{m,n} P E_{kin}^{1/2}}$ where $P=k_\rho |F|^2 dk_\rho dk_z$.}
N is the probability that the wave function is inside the nucleus at a given 
time T.
\label{fig-1}       % Give a unique label
\end{figure*}

\begin{figure*}[!htbp]
%\begin{figure}[h]
\centering
\includegraphics[width=0.80\textwidth,clip]{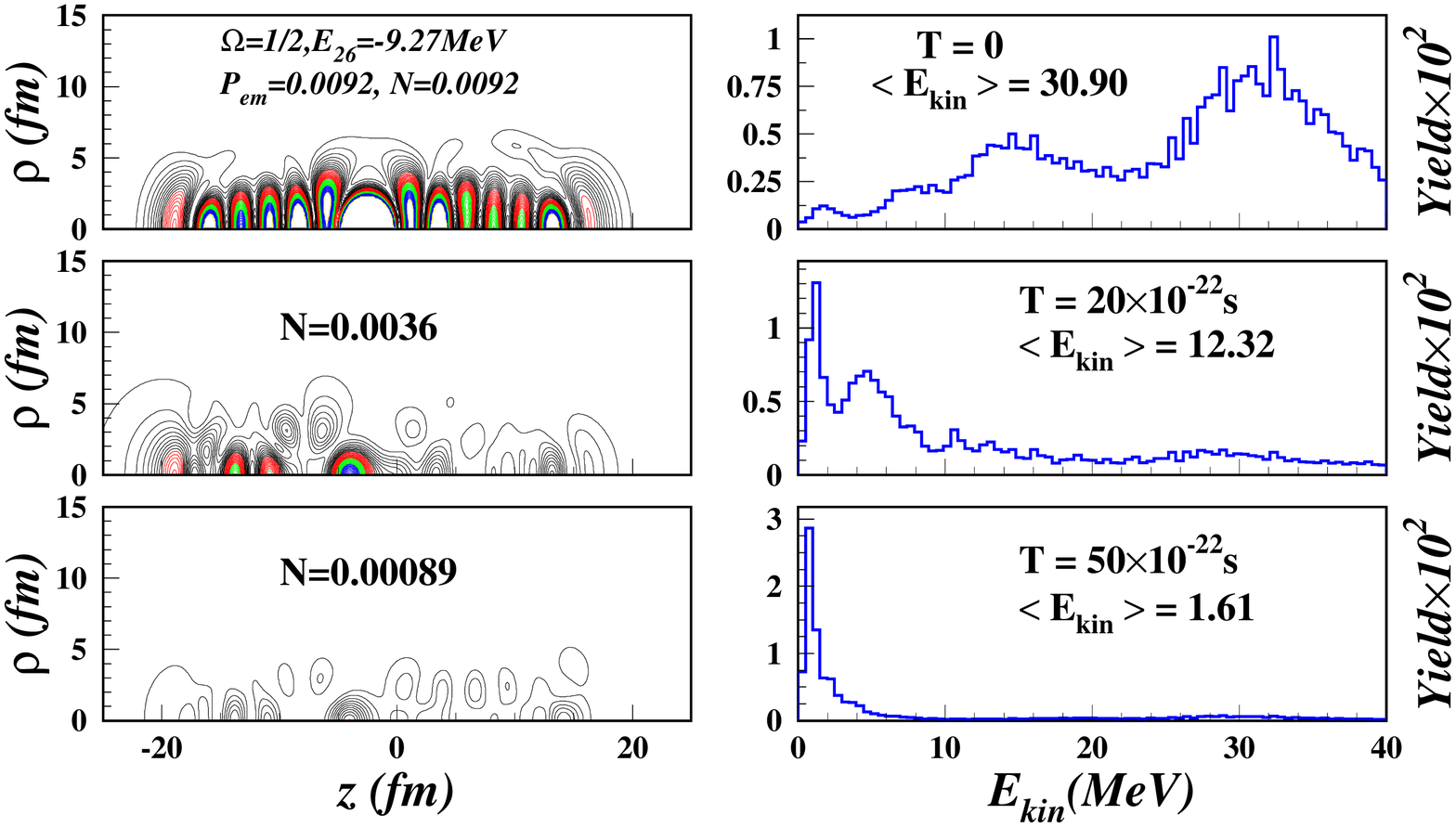}
\caption{The same as in Fig.1 but for the unbound WF$_{26}$.}
\label{fig-2}       % Give a unique label
\end{figure*}

\begin{figure*}[!htbp]
%\begin{figure}[h]
\centering
\includegraphics[width=0.80\textwidth,clip]{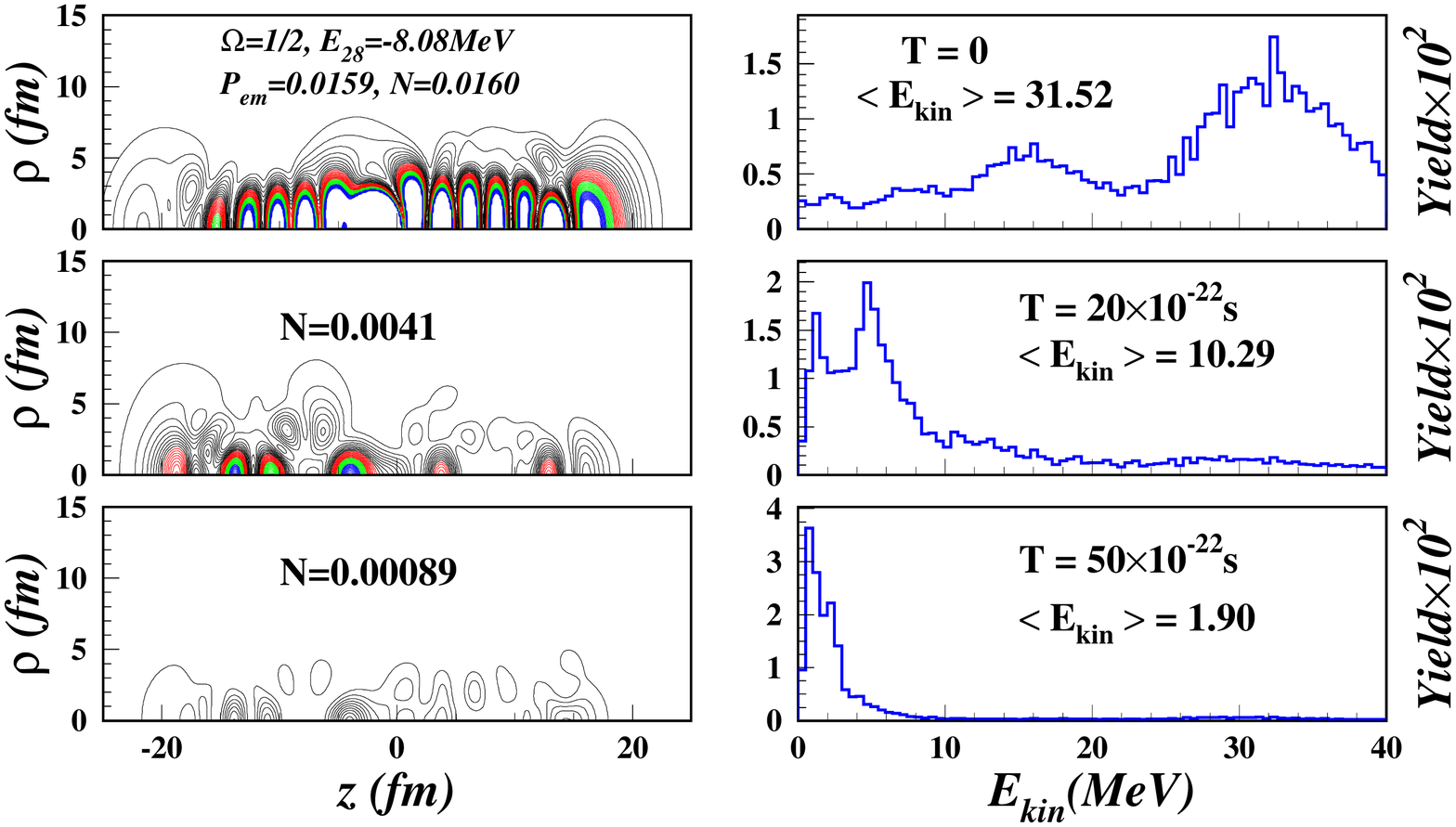}
\caption{The same as in Fig.1 but for the unbound WF$_{28}$.}
\label{fig-3}       % Give a unique label
\end{figure*}

\begin{figure*}[!htbp]
\centering
\includegraphics[width=0.80\textwidth,clip]{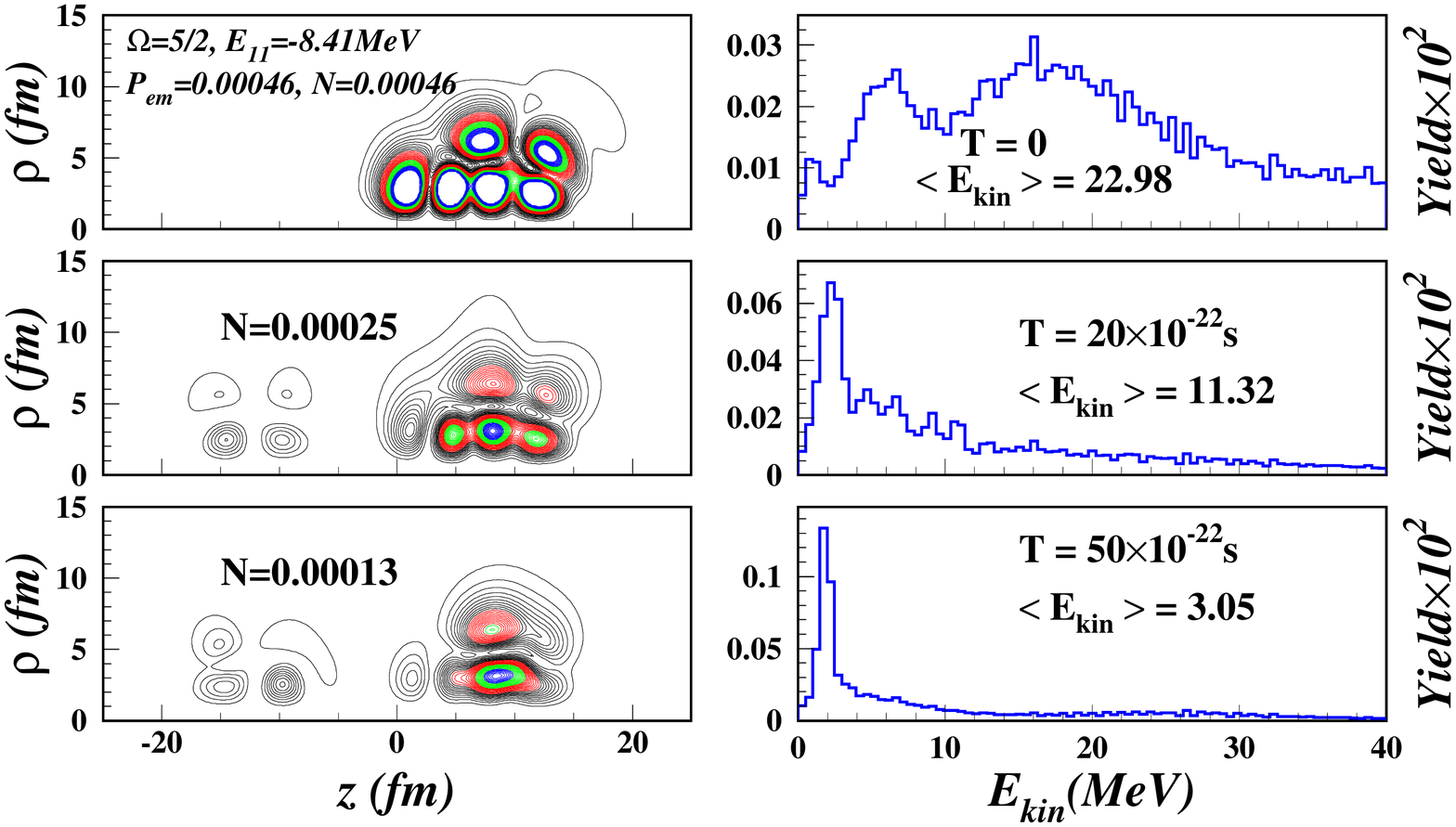}
\caption{Square modulus of the unbound WF$_{11}^{5/2}$ (left column) and energy
distribution (right column) at
different times T. The projection of the angular momentum on the fission axis
of this wave function is $\Omega$=5/2.
The wavefunctions at $T=0$ and $50\times 10^{-22}$ sec are represented relative
to that at $T=20\times 10^{-22}$ sec.
The values on the ordinates of the histograms are $P_{11}(E_{kin})$
probabilities multiplied by $10^2$.
$E_{kin}^{mean}= \frac{\sum_{m,n} E_{kin} P E_{kin}^{1/2}}
{\sum_{m,n} P E_{kin}^{1/2}}$ where $P=k_\rho |F|^2 dk_\rho dk_z$.}
N is the probability that the wave function is inside the nucleus at a given
time T.
\label{fig-4}       % Give a unique label
\end{figure*}

\begin{figure*}[!htbp]
%\begin{figure}[h]
\centering
\includegraphics[width=0.80\textwidth,clip]{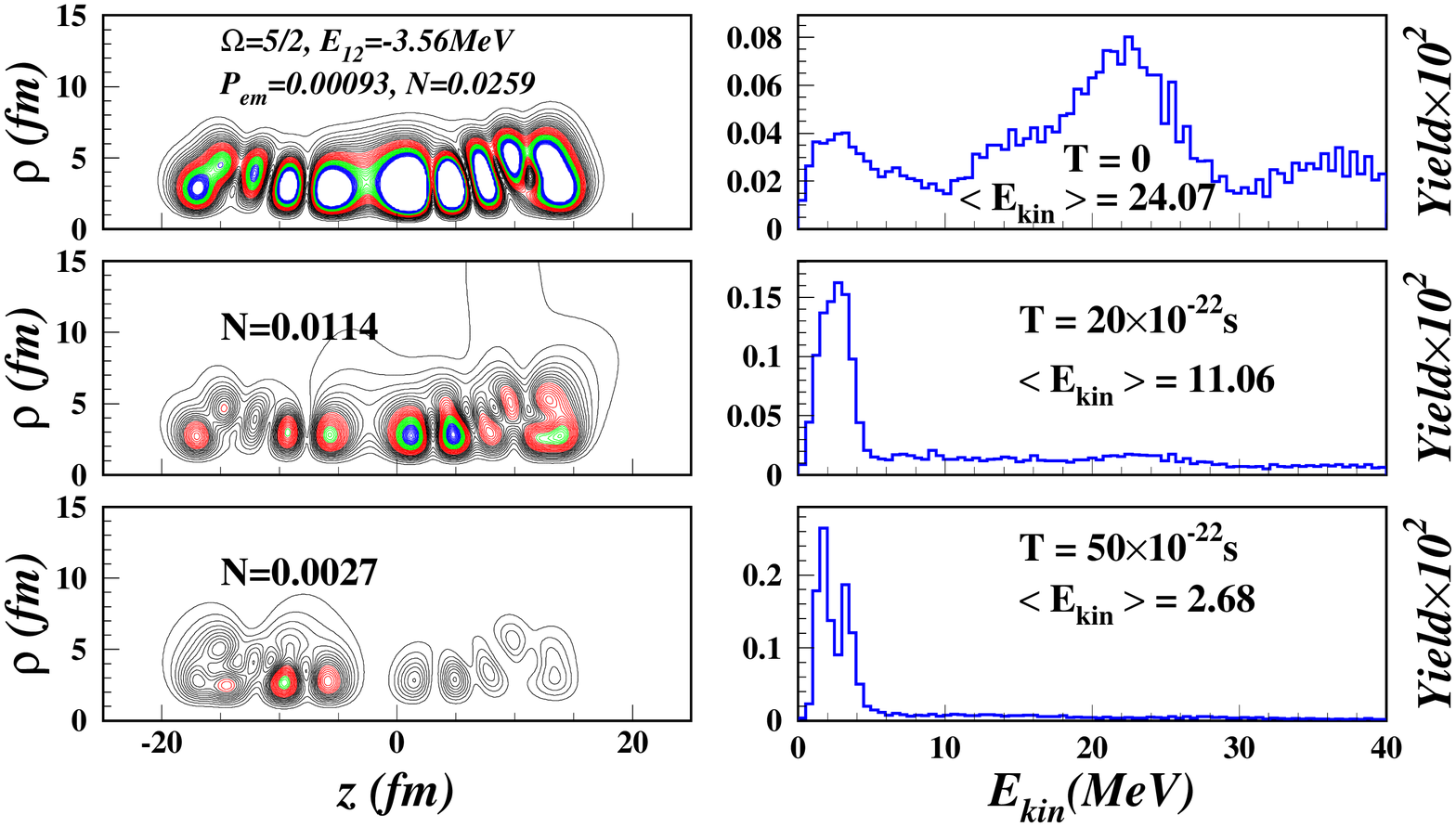}
\caption{The same as in Fig.4 but for the unbound WF$_{12}^{5/2}$.}
\label{fig-5}       % Give a unique label
\end{figure*}

\begin{figure*}[!htbp]
%\begin{figure}[h]
\centering
\includegraphics[width=0.80\textwidth,clip]{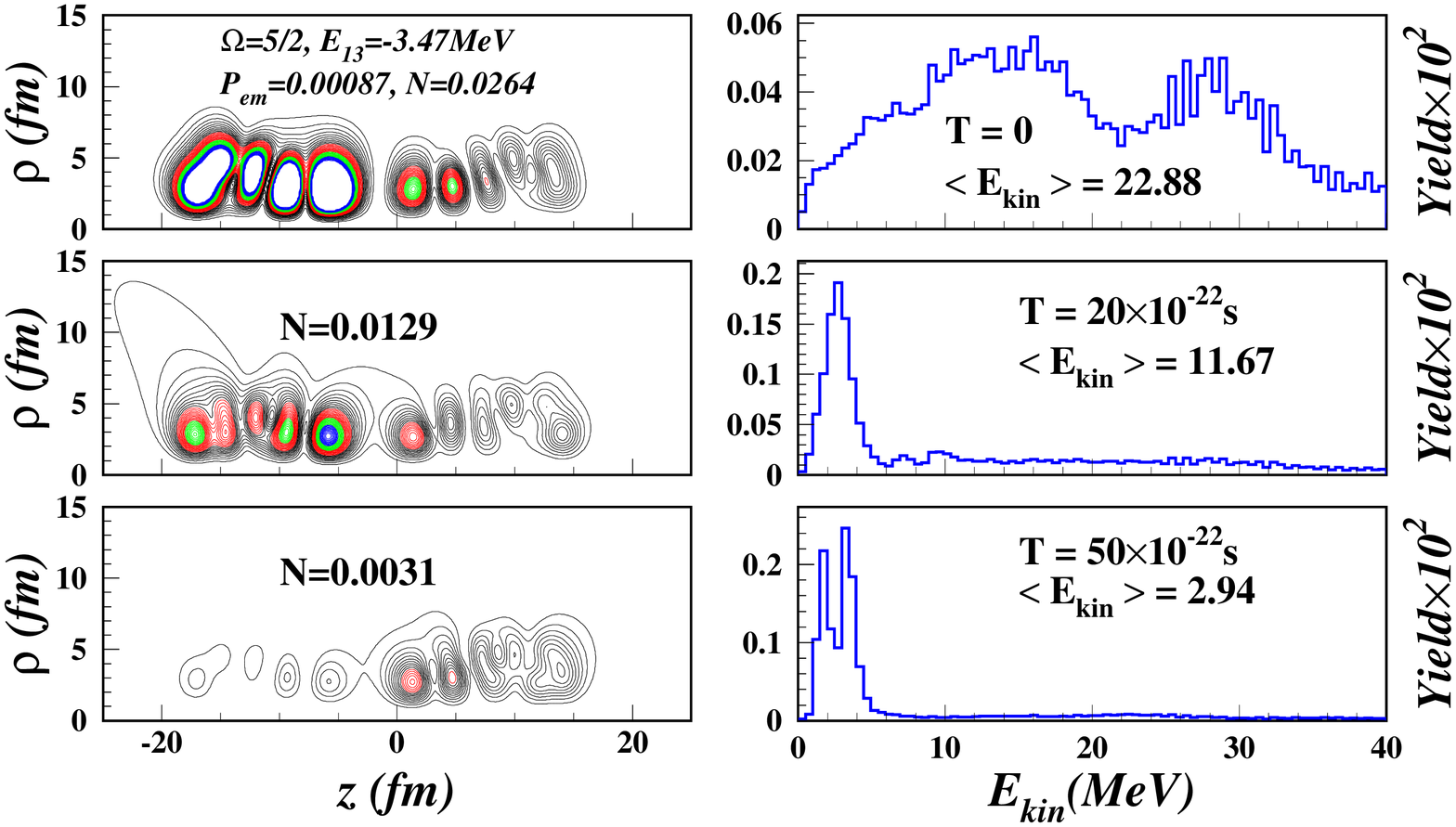}
\caption{The same as in Fig.4 but for the unbound WF$_{13}^{5/2}$.}
\label{fig-6}       % Give a unique label
\end{figure*}

\begin{figure*}[!htbp]
\centering
\includegraphics[width=0.80\textwidth,clip]{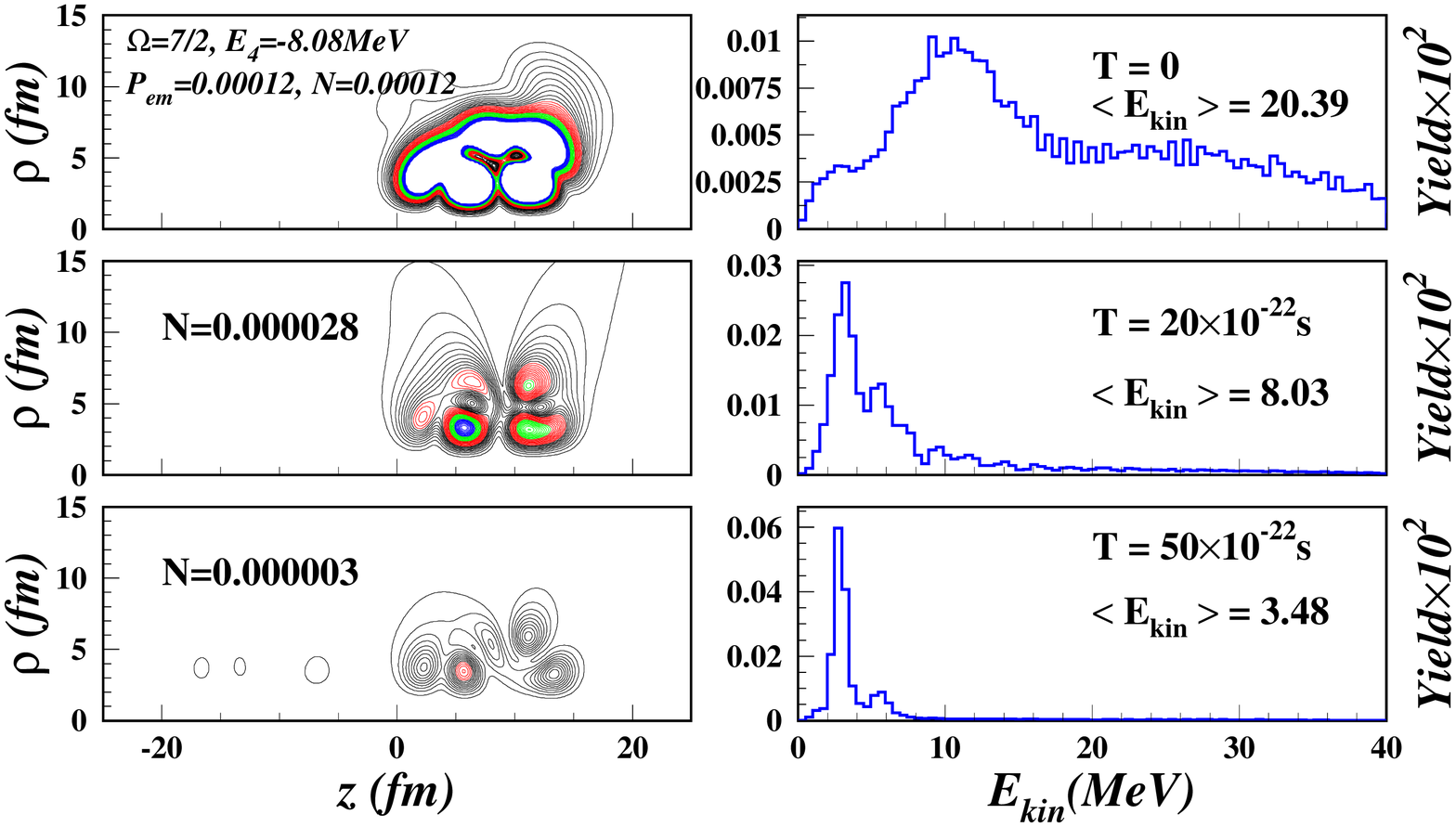}
\caption{Square modulus of the unbound WF$_{04}^{7/2}$ (left column) and energy
distribution (right column) at
different times T. The projection of the angular momentum on the fission axis
of this wave function is $\Omega$=7/2.
The wavefunctions at $T=0$ and $50\times 10^{-22}$ sec are represented relative
to that at $T=20\times 10^{-22}$ sec.
The values on the ordinates of the histograms are $P_{04}(E_{kin})$
probabilities multiplied by $10^2$.
$E_{kin}^{mean}= \frac{\sum_{m,n} E_{kin} P E_{kin}^{1/2}}
{\sum_{m,n} P E_{kin}^{1/2}}$ where $P=k_\rho |F|^2 dk_\rho dk_z$.}
N is the probability that the wave function is inside the nucleus at a given
time T.
\label{fig-7}       % Give a unique label
\end{figure*}

\begin{figure*}[!htbp]
%\begin{figure}[h]
\centering
\includegraphics[width=0.80\textwidth,clip]{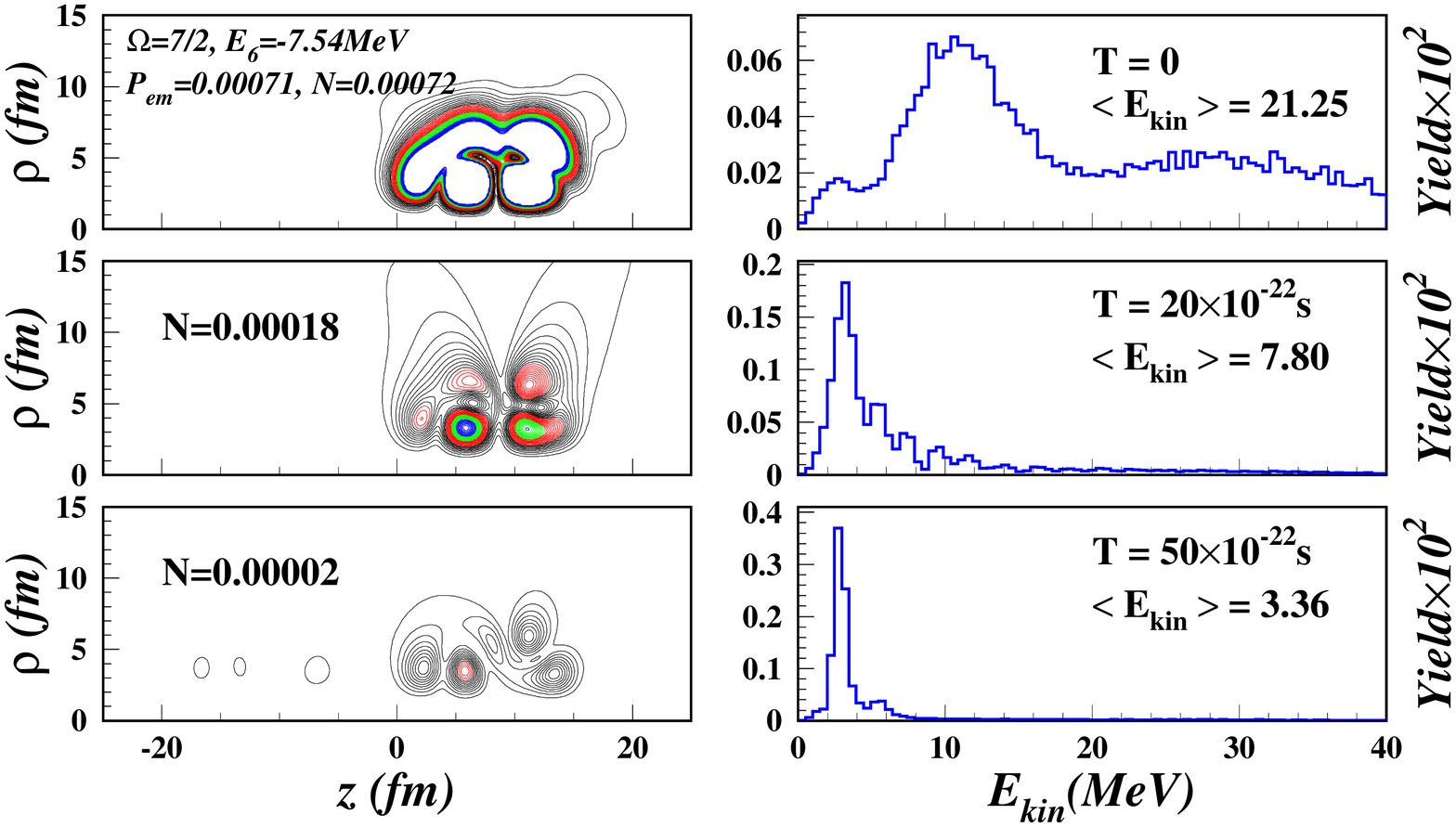}
\caption{The same as in Fig.7 but for the unbound WF$_{6}^{7/2}$.}
\label{fig-8}       % Give a unique label
\end{figure*}

The initial wave packets are given by Eq.(4). 
At $T$=0, i.e. immediately after scission, the released neutron populates bound 
states in the continuum and it is mainly localized in the neck region since it 
is there that the potential changes mostly. 
The kinetic energy of the unbound neutron can reach values as high as  
the potential depth $V_0$ which is $40.2$ MeV in our case. The average value 
is however lower (around $30$ MeV) due a large diffuse surface and tails of the 
wave functions that penetrate into the potential wall.    
One notices that with increasing time ($T$ = $20$ and $50\times 10^{-22}$ sec) 
the amplitude of the wavefunctions  
diminishes, showing that the neutron is leaving the nucleus. 
At the same time, the $E_{kin}$-distribution is shifted to lower values, 
reflecting the fact that the neutrons are less and less present inside the 
potential well. 

At very large times the 
neutron should be completely emitted. One sees that, due to
numerical limitations, we cannot reach this situation: even at $T_{max}$ the 
neutron has still 10$\%$ probability to be inside the fragments.
If we calculate longer, the part of the wave packet that is reflected on the 
boundary of the spatial grid returns inside the nucleus affecting the energy 
spectrum. $T_{max}$ is therefore related to the size of the ($\rho,z$) grid 
used.  
Since the above mentioned probability is small, one can consider that at 
$T=50\times 10^{-22}$ sec 
the calculated $E_{kin}$-distribution represents well the emitted neutron. 
These single spectra are characterized by a peak at low energies (below $2$ 
MeV) plus a short tail towards higher energies.

So far we have analyzed energy distributions for wave functions with 
$\Omega$=1/2 which correspond to orbital angular momentum projections
$\Lambda$= 0 or 1. In most cases $|f_1|^2 >> |f_2|^2$, see Eq.(2), 
so there is practically no centrifugal barrier.   

However, for larger $\Omega$ values, the centrifugal potential,
$\Lambda_{1,2}^2 /\rho^2$, is expected to 
play a role. In the figures 4 to 6 are shown emitted wave function with 
$\Omega$=5/2 ($\Lambda$=2 or 3) and indices 11, 12 and 13 and the corresponding 
kinetic energy histograms. As compared with the previous case:  

a) At $T$=0 the square moduli of the unbound wave functions,
are displaced from the z axis where the centrifugal potential has a 
maximum. Of course this comes from the same feature of the total wave functions.
For this reason they are less present in the neck region and contribute less to 
the scission neutron multiplicity.     

b) At $T$=0 the spectrum is shifted towards lower values since the kinetic 
energy 
is reduced by the centrifugal potential. As a result the average kinetic 
energy is smaller ($\approx$23 MeV). 
At $T=50\times 10^{-22}$ sec the average kinetic energy is larger 
($\approx$3 MeV) since 
the centrifugal potential is now transformed into kinetic energy. 

Figs. 7 and 8 show the time evolution of the wave packets and of the kinetic 
energy 
histograms for states (indices 4 and 6) corresponding to an even higher 
$\Omega$ value (7/2). As expected, $<E_{kin}>$ becomes even lower 
($\approx$21 MeV) 
at $T$=0 and even larger ($\approx$3.5 MeV) at $T=50\times10^{-22}$sec.
An interesting feature of the individual spectra at high $\Omega$ values is 
the existence of a peak at low energies in the initial spectrum, the intensity 
of which increases with 
time. It reflects the fact that the wave functions, being located at the 
nuclear surface, spread outside the fragments even at $T$=0.

\section{Scission neutron spectrum}
\label{sec-5}

To obtain the whole kinetic energy spectrum for a fixed mass asymmetry,
one has to sum the single spectra over all occupied states 
and all $\Omega$ value. 

\begin{figure*}[!htbp]
\centering
%\sidecaption
\includegraphics[width=0.59\textwidth]{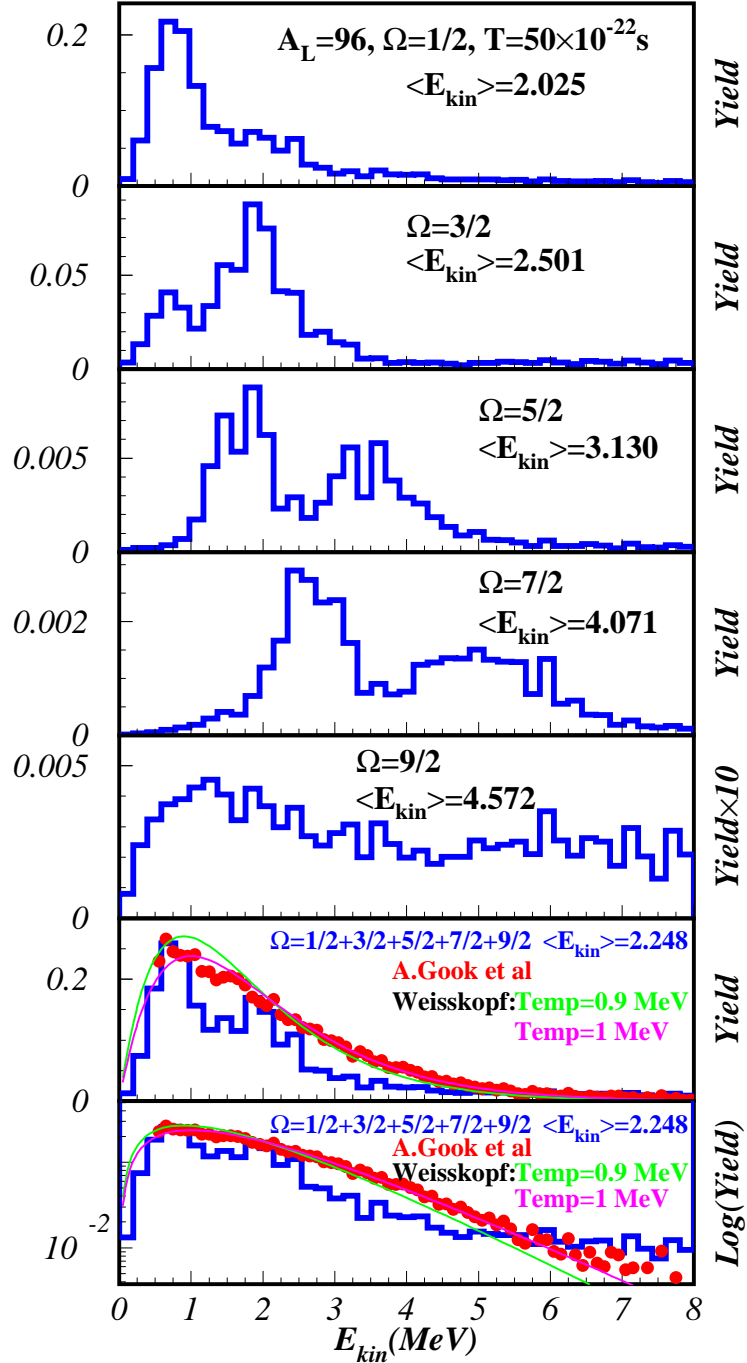}
\caption{Kinetic energy distributions at $T=50\times 10^{-22}$ sec for 
sub-states defined by the quantum number $\Omega$. 
In the two lowest frames the distribution calculated with all 
neutron states is presented together with recent experimental results 
\cite{Alf} from the reaction $^{235}U(n_{th},f)$. 
Two typical evaporation spectra \cite{Weissk}
characterized by nuclear temperatures $Temp = 1.0$ and $0.9$ MeV are also 
plotted for comparison.  The evaporation spectra were
normalized to the data but there is an arbitrary normalization between the
data and the calculation.  
($Yield = \sum P_i(E_{kin})\times v_i^2$).} 
\label{fig-9}       % Give a unique label
\end{figure*}

The upper five frames of Fig. \ref{fig-9} show kinetic energy spectra for 
$\Omega$=1/2, $\Omega$=3/2, $\Omega$=5/2, $\Omega$=7/2 and 
$\Omega$=9/2 respectively. 
The kinetic energy increases with increasing $\Omega$ due to the 
the centrifugal term. 
Note however that $\Omega=1/2$ gives the dominant contribution (65$\%$).

In the 6th frame, the total spectrum (summed over the five $\Omega$
values) is shown. 
It presents a maximum around $0.7$ MeV and an 
exponentially decreasing tail till $8$ MeV in qualitative agreement with 
the measured spectrum \cite{KoHa} of all prompt fission neutrons (PFN). 
For comparison, we added recent data \cite{Alf} obtained for the same 
constraint on mass asymmetry ($A_L$ = 96). 
One notices that both the data and the calculation are not smooth.
The oscillations in the data are statistically significant. 
The calculated distribution is not smooth since it consists of a finite 
weighted sum of individual contributions with different mean values and widths. 
The number of non-negligible terms is only 35, distributed among the 
$\Omega$ values as following: 21 for 1/2, 8 for 3/2, 4 for 5/2 and 2 for 7/2.
Hence less than half of the total number of the neutrons in $^{236}U$ 
contribute significantly to the scission neutron spectrum. 

However the data do not oscillate as much as the calculations.
One reason is that the data are affected by a finite energy resolution.
If we convolute the theoretical spectrum
with a Gaussian resolution function, the amplitude of its oscillations will
decrease. The other reason is that our model (as any model) contains
approximations and numerical limitations.

Two typical evaporation spectra \cite{Weissk}, $E \exp (-E/Temp)$,
for nuclear temperatures $Temp = 1.0$ and $0.9$ MeV are also ploted.
We stress that, in this case ($A_L$ = 96), each fragment
evaporates about one neutron on the average and the Weisskopf formula
should work.
These evaporation spectra follow quite well the general trend of the recent 
data except at very low and very high energies. 
$Temp=0.9$ MeV reproduces better the drop at low energies while $Temp=1.0$ MeV 
the tail at high energies. Evidently, none of them exhibit oscillations.    

In the lowest frame of Fig. 9 the same comparison is shown in lin-log scale 
to unveil hidden differences at $E_{kin} >$ 5 MeV. 
One can see that, in contrast to the EVN, the SN can reproduce the high 
energy tail of the PFN spectrum. 
This inability of the evaporation hypothesis to account for high energy PFN  
has been already discussed in Ref.\cite{Kornilov}.

%\begin{figure}[h]

\section{Summary}
\label{sec-6}

The dynamical scission model \cite{RCNPA13} is used to calculate SN kinetic 
energy spectra, at different intervals of time after scission,
for the fission of $^{236}U$ into the most probable mass division ($A_L$=96). 
The evolution of the wave packets $|\Psi^i_{em}|^2$ (representing  the 
neutrons released during scission) and of their kinetic energy, $E_{kin}$, 
distributions 
reflects the process of separation of the scission neutrons from the nascent 
fission fragments.

The spectrum at the largest time we were able to attain numerically (i.e., 
$T_{max} = 5\times 10^{-21}$ sec)  is compared with recent measurements 
obtained with high statistics and resolution \cite{Alf} in the reaction 
$^{235}U(n_{th},f)$ for the same mass division.
As in the case of the PFN angular distribution \cite{PLB,CRT-ND16}, both 
hypotheses (evaporation from fully accelerated fragments and dynamical 
emission at scission) explain satisfactorily the general features of the 
measured spectrum. This difficulty to distinguish experimentally between 
two completely oposite theoretical assumptions is puzzling.

There is however a detail that makes the results of the two hypotheses 
slightly different: the evaporation spectrum is smooth while the SN spectrum 
presents structures due to the finite number of neutrons that contribute.

In spite of computational limitations 
(not large enough $(\rho,z)$ grid and not long enough time evolution),
a more quantitative agreement could be forseen by including the simultaneous 
separation of the fragments after scission and by taking into account the 
re-absorption of the unbound neutrons by the imaginary potential of the nascent 
fragments. Such calculations are in progress.

\section*{Acknowledgements}
This work was done in the frame of the projects: 
PN-III-P4-ID-PCE-2016-0649 (contract nr. 194/2017), 
PCE-2016-0014 (contract nr.7/2017) and EC EURATOM FP7 project CHANDA.

\appendix*

\section{Total neutron wave packets and their emitted parts at $\alpha_f$}

As stated in Sec 2, after a diabatic transition at scission, all 
neutrons are represented by expansions in the set of eigenstates of the 
nuclear configuration $\alpha_f$. At the higher end, these wave packets are 
built on states in the continuum which can therefore leave the nucleus.
In the dynamical scission model these small parts, defined by Eq. (4), are 
the scission neutrons. From pedagogical point of view it is useful to 
vizualize and understand the differences between the total wave packet and 
its tiny unbound tail. 
 
\begin{figure*}[!htbp]
\centering
\includegraphics[width=0.60\textwidth,clip]{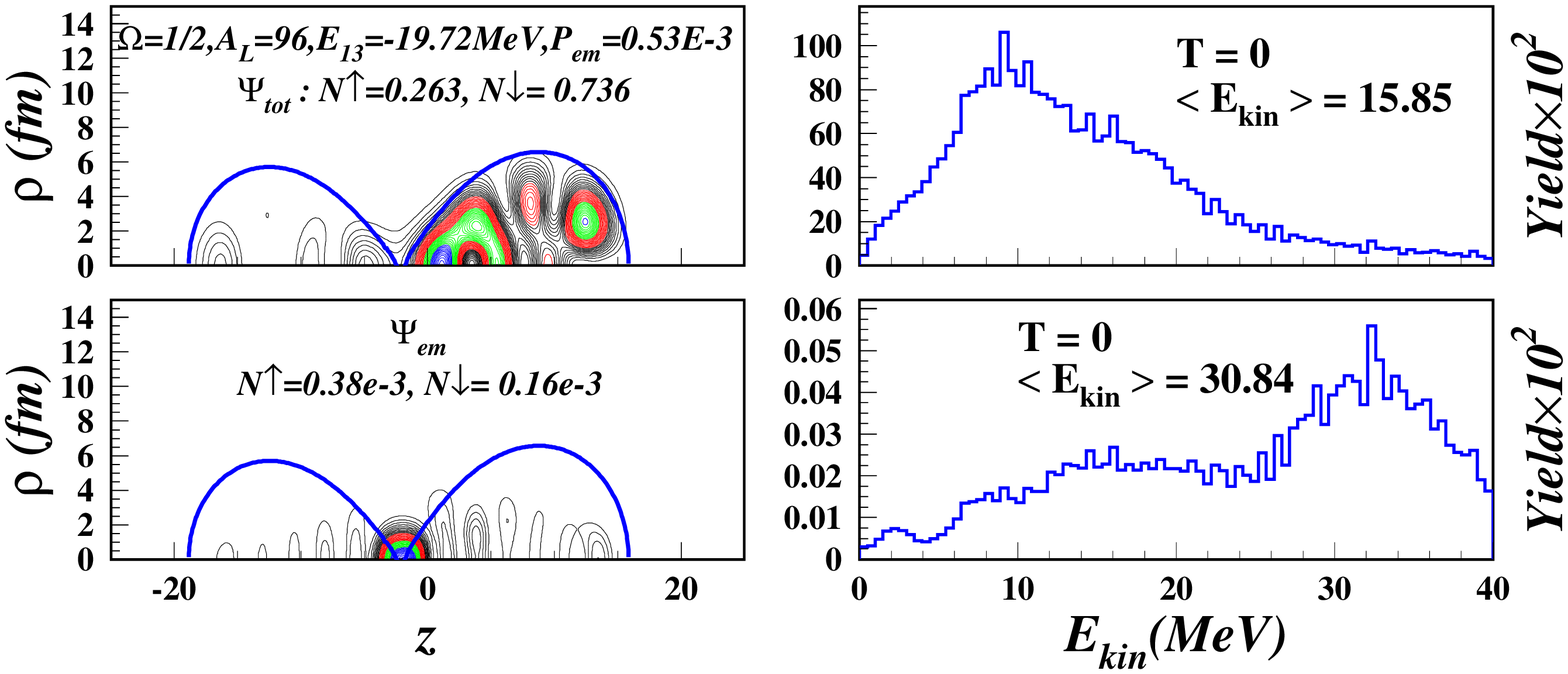}
\caption{Square modulus of the total and emitted WF$_{13}$ at 
$\Omega=1/2$ immediately after scission (left column) and the corresponding 
energy distributions (right column). N$\uparrow$ and N$\downarrow$ are the 
square moduli of the spin-up and spin-down components.
}
\label{fig-10}       % Give a unique label
\end{figure*}

\begin{figure*}[!htbp]
\centering
\includegraphics[width=0.60\textwidth,clip]{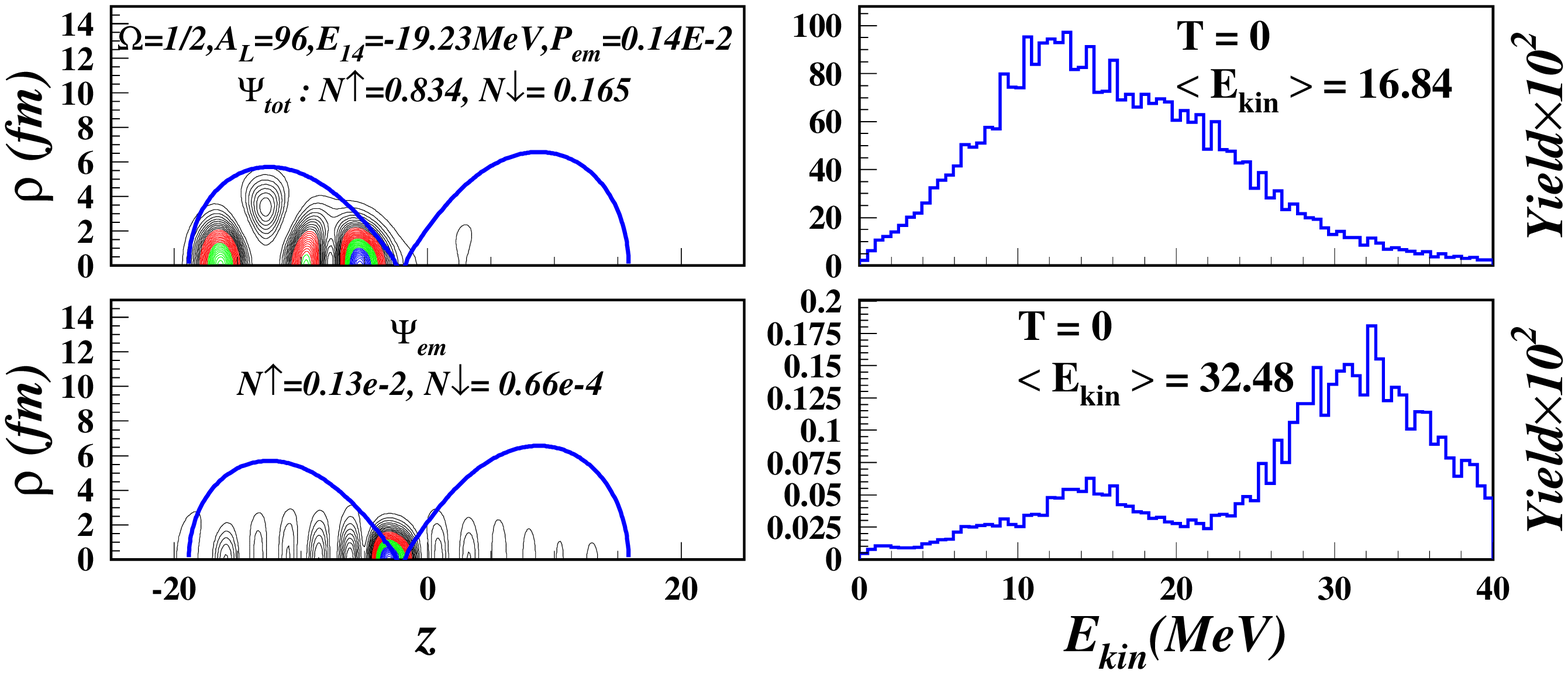}
\caption{The same as in Fig.10 but for WF$_{14}$ at $\Omega=1/2$}
\label{fig-11}       % Give a unique label
\end{figure*}

\begin{figure*}[!htbp]
\centering
\includegraphics[width=0.60\textwidth,clip]{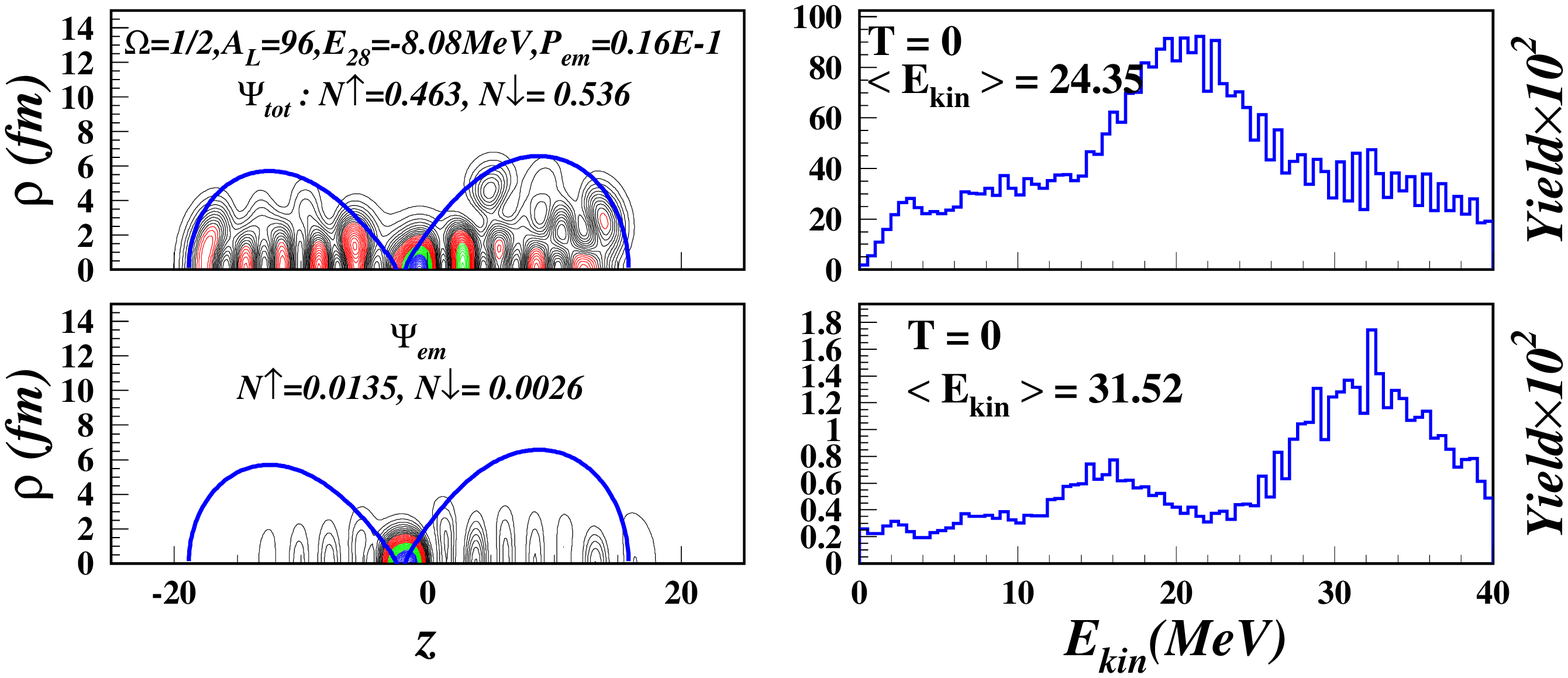}
\caption{The same as in Fig 10 but for WF$_{28}$ at $\Omega=1/2$}
\label{fig-12}       % Give a unique label
\end{figure*}

In the figures \ref{fig-10} - 
\ref{fig-12} are shown three wave packets corresponding 
to $\Omega=1/2$. For states with low energies the total 
wave functions (WF$_{13}$ and WF$_{14}$) are confined in one of the 
fragments (light or heavy).  
The emitted wave functions are concentrated in the neck region where the 
coupling to the changing potential is the strongest. 
For states with high energies, the total wave functions 
are localized in both fragments (like WF$_{28}$ in Fig.\ref{fig-12}).
Equipotential lines corresponding to $V_0/2$
are also plotted to represent the fragments immediately after scission.
As expected, the part of the wave packet that is emitted has higher average 
energy and more nodes (a larger quantum number).
 
\begin{figure*}[!htbp]
\centering
\includegraphics[width=0.60\textwidth,clip]{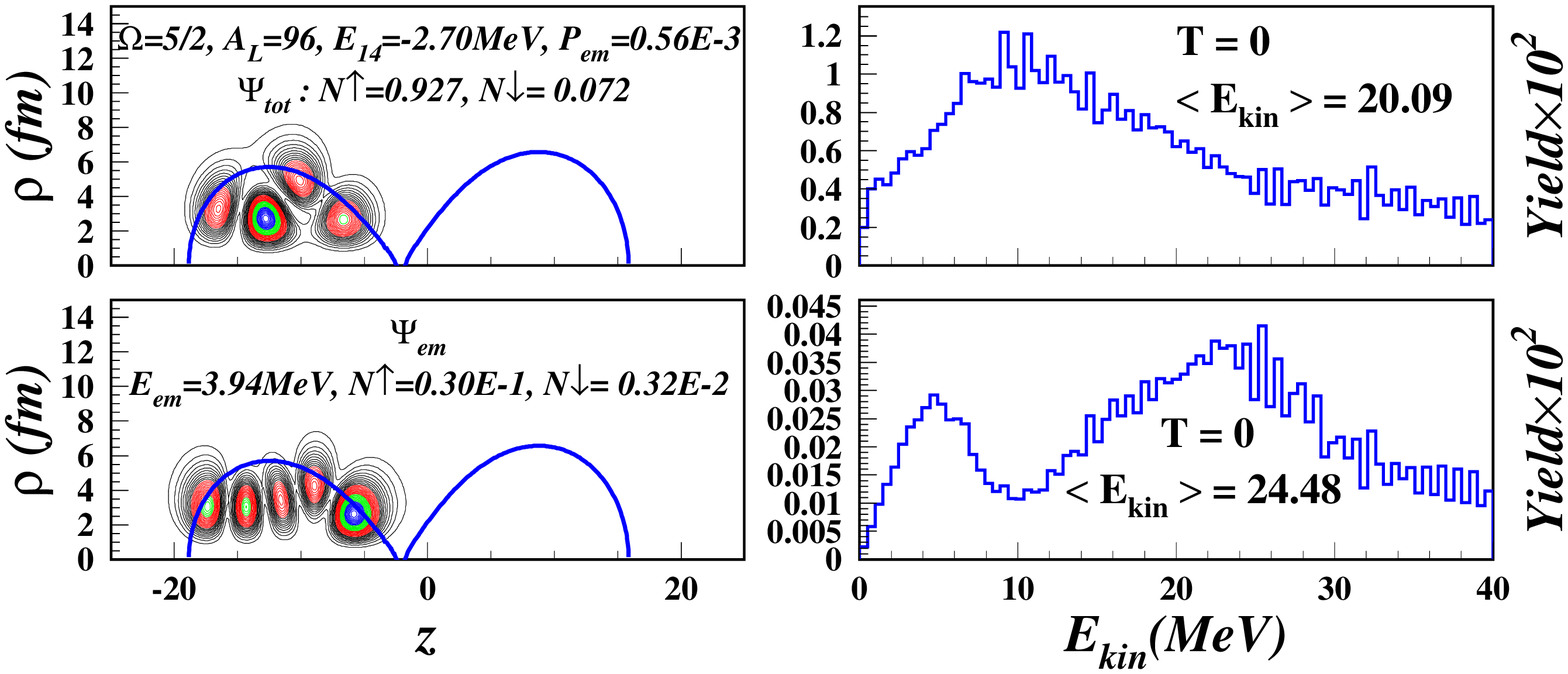}
\caption{Square modulus of the total and emitted WF$_{14}$ at
$\Omega=5/2$ immediately after scission (left column) and the corresponding
energy distributions (right column)}
\label{fig-13}       % Give a unique label
\end{figure*}

\begin{figure*}[!htbp]
\centering
\includegraphics[width=0.60\textwidth,clip]{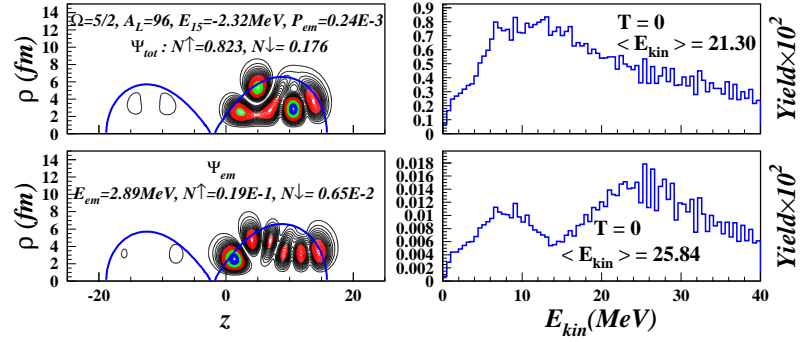}
\caption{The same as in Fig. 13 but for WF$_{15}$ at $\Omega=5/2$}
\label{fig-14}       % Give a unique label
\end{figure*}

In the figures \ref{fig-13} - \ref{fig-14} are shown two wave packets at 
$\Omega=5/2$. The total wave functions are restricted to only one of the 
fragments.
One can see the effect of the centrifugal potential: the wave functions are 
shifted with respect to the $\rho$ axis.   
For this reason they cannot be present in the neck region where the potential 
changes mostly. Their contribution to scission neutron multiplicity is 
therefore reduced. 

\end{document}